# Multifractality of complex networks is also due to geometry. The Geometric SandBox algorithm.

Rafał Rak[1], Ewa Rak[2]

October 13, 2024

[1]College of Natural Sciences, Institute of Physics, University of Rzeszów, Pigonia 1, 35-310 Rzeszów, Poland

[1]College of Natural Sciences, Institute of Mathematics, University of Rzeszów, Pigonia 1, 35-310 Rzeszów, Poland

**Abstract**

Over the past three decades, describing the reality surrounding us using the language of complex networks has become very useful and therefore popular. One of the most important features, especially of real networks, is their complexity, which often manifests itself in a fractal or even multifractal structure. As a generalisation of fractal analysis, multifractal analysis of complex networks is a useful tool for the identification and quantitative description of the spatial hierarchy of both theoretical and numerical fractal patterns. Nowadays, there are many methods of multifractal analysis. However, all these methods take into account only the fact of connection between nodes (and eventually the weight of edges) and do not take into account the real positions (coordinates) of nodes in space. However, intuition suggests that the geometry of network nodes' position should have a significant impact on its true fractal structure. Many networks identified in nature (e.g. air connection networks, energy networks, social networks, mountain ridge networks, networks of neurones in the brain, street networks) have their own often unique and characteristic geometry, which is not taken into account in the identification process of multifractality in commonly used methods. In this paper, we propose a multifractal network analysis method that takes into account both connections between nodes and the location coordinates of nodes (network geometry). We show the results for different geometrical variants of the same network and reveal that this method, contrary to the commonly used method, is sensitive to changes in network geometry. We carry out tests for synthetic as well as for real-world networks.

# 1   Introduction

Mathematical modelling of networks dates back to the late 1950s, when Erdős and Rényi initiated the field of random graphs [1] and scientists began to develop it. Mathematically, a network is a representation of a real complex system and is defined as a collection of nodes (vertices) and links (edges) between pairs of nodes. Complex networks have naturally become a convenient tool for studying complex systems where many elements (nodes) are observed and connected with each other by a certain interaction (edge). It turns out that one of the greatest challenges in the world of science is a precise and complete description of complex systems, and network research has become an important and indispensable element of this process. Due to their usefulness in studying real-world complex systems, the study of complex networks and multifractal analysis has been developed in many fields such as mathematics, physics, and chemistry [2, 3, 4], biological systems [5, 6, 7, 8, 9], economics [10, 11], computer science [12, 13, 14, 51, 52, 17], language and sociology [18, 19, 21, 22, 23], or geology [24].

Most real networks show interesting topological features and inter-node connections are complex nature. Often, the study of their characteristics boils down to the analysis of: assortativeness between vertices, clustering coefficient, degree distribution, reciprocity, centrality, or shortest paths. It seems that a useful and promising tool for identifying the possible complexity of the network structure is fractal analysis, which is one of the most dynamically developing topics in the last two decades. Moreover, it was noticed that complex structures can be more effectively characterised in the process of multifractal analysis, where a function of dimensions is replaced by a non-trivial resultant of single fractal dimensions. One of the most important results in physics was the description of fractal geometry by Mandelbrot [25]. While these beautiful geometric fractal structures apply to structures in physical space, new forms of fractality have been observed in networks, which is the result of complex interactions between nodes. Nowadays, it is assumed that a network's fractality is the result of its individual or resultant properties and mechanisms such as: self-similarity, growth phenomenon, small-world phenomenon, scale-free degree distribution, self-organization [26, 27, 28, 29, 30, 31, 32, 33].

There are two main types of methods for identifying (multi- )fractal properties of complex networks: the cluster-growing method and the box-covering method. Classical box-covering algorithms are focused on solving the problem of how to cover the whole network with the minimum boxes, which is related to the family of NP-hard problem (non-deterministic polynomial-time) [35]. This family of algorithms includes the compact-box-burning [35], maximum excluded-mass-burning [36], the fixed-size box-counting [37, 38] and random sequential box-covering [39, 40]. Cluster-growing methods are an alternative to the traditional box-counting(covering) method. They are based on the method proposed in [41, 42]. In simplification, this method resembles the sandbox and focuses on the scaling of the masses versus the size through sandboxes growing from randomly selected centres. To avoid ambiguity in the results due to the randomness of the method, calculations are repeated for many nodes(seeds), and then averaged. The extension of the sandbox method for multifractal analysis of complex networks was proposed by Liu et al. [43], which supported the fact that the sandbox algorithm is more accurate than box-counting algorithms to calculate multifractal parameters. The method has also been improved and extended to weighted networks [44, 45, 46, 47].

Characterising the complexity of the network is very important, e.g. because the functions of the network that represents a given system are the resultant of its structure. For example, the structure (geometry; relative positions) of the airport connection network affects its efficiency; the structure of the network of synaptic connections in the brain affects its efficiency; position of atoms in materials affects the chemical and mechanical properties of the materials. Therefore, it should be emphasised that both the box-covering method and the sanbox method do not take into account the network parameters (e.g. the fact that nodes are connected) and their coordinates in space and mutual positions in relation to each other (geometry). In order to meet the above facts, in this paper we propose a method of multifractal analysis of complex networks based on both box-counting and sandbox methods.

## 2 Sandbox algorithms

### 2.1 The standard sandbox algorithm

The idea and main steps of the existing Sandbox algorithm (SB) for multifractal analysis of network $Q$ can be described as follows:

(1) The set of the radius $r$ ($1 \leq r \leq d$) of the sandbox is determined, where $d$ denotes the diameter of the network $Q$.

(2) Randomly choose a node set $C(r)$ as the centres of sandboxes.

(3) Count the number of nodes $M_i(r)$ covered by the sandbox with the center node $i \in C(r)$ and radius $r$. This process for one of the centre node $i$ and $r = \{2, 7, 8\}$ is illustrated in Fig.1. It should be emphasized that $r$ is not a geometric radius but a measure of the propagation over the closest neighbour nodes of the $Q$ network.

(4) For each scaling parameter $q \neq 1$ calculate the average $\langle [M_i(r)]^{q-1} \rangle$ over all sandboxes of radius $r$;

(5) For a fixed set of radii selected from $1 \leq r \leq d$ repeat steps (2)–(4);

(6) If linear dependence between $\ln \left( \langle [M_i(r)]^{q-1} \rangle \right) /(q-1)$ and $\ln(r/d)$ is observed then we can talk about the (multi)fractal nature of the investigated network $Q$ and calculate the reliable generalized fractal dimension $D_q$ and the singularity spectrum $f(\alpha)$ as follows:

$$D_q = \lim_{r \to 0} \frac{\ln \langle [M_i(r)]^{q-1} \rangle}{\ln(r/d)} \frac{1}{q-1}, \quad q \in \text{Re}, \quad q \neq 1, \tag{1}$$

$$\tau_q = (q-1)D_q; \quad \alpha = \frac{d}{dq}\tau_q \text{ and } f(\alpha) = q\alpha - \tau_q, \tag{2}$$

where $\alpha$ is called the singularity (Hölder) exponent. The wealth of multifractality present in analyzed $Q$ network can be defined both as a spread of the generalized fractal dimension $D_q$: $\Delta D_q = D_{q_{min}} - D_{q_{max}}$ or as a width of multrifractal spectrum $f(\alpha)$: $\Delta \alpha = \alpha_{max}(q_{min}) - \alpha_{min}(q_{max})$, where $q_{min}$ and $q_{max}$ are respectively the minimal and the maximal value of the deformation parameter $q$.

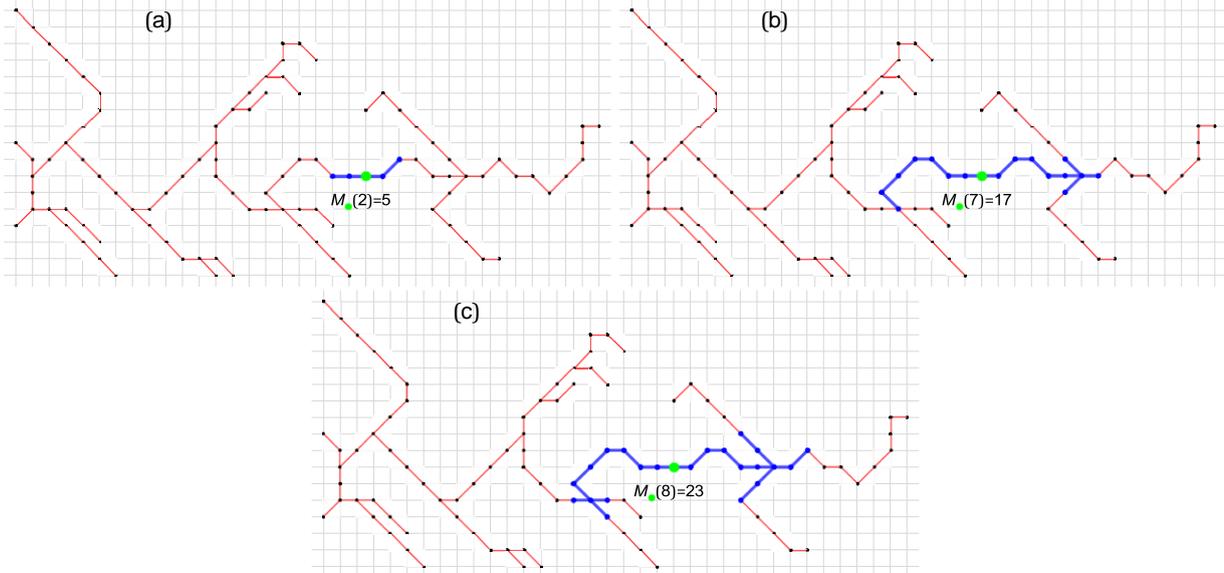

Figure 1: The standard sandbox algorithm. The number of nodes $M(r)$ covered by the sandbox with the same center node for three different radius $r = \{2, 7, 8\}$.

## 2.2 The geometric sandbox algorithm

Let us consider the same network in different geometrical representations where nodes have the same structure of connections but are distributed differently in relation to each other in the plain (Fig.2, Fig.3). Changing the position of the nodes changes the length of the edges, but does not change the node's degree. Without taking into account the weights between nodes, from the network point of view, these networks are identical – for each of them, the result of the standard SB algorithm is identical. However, our intuition tells us that these networks may (and even should have) different characteristics of complexity. Of course, there is a sandbox version of the algorithm that takes into account the weights of connections between nodes where the weight can be e.g. the distance between two neighbouring nodes. However, this is just an interaction between neighbours, and more global peer-to-peer interactions are not considered. In order to take into account both the geometric position of the nodes and their network properties in the multifractal analysis, we propose a modified SB algorithm - the Geometric Sandbox algorithm (GSB). The next steps of the GSB algorithm are as follows:

(1) The set of the radius of circles $R$ ($0 < R \leq P$) of the sandbox is determined, where $P$ denotes the geometric diameter (the distance between the most distant nodes) of the network $Q$. In practise, however, we choose the smaller $R$ ($0 < R \leq P/5$). Unlike the standard SB, here the radius $R$ is not, in general, an integer number (expressed as multiples of the edge), but is a geometric distance (a real number).

(2) Randomly choose a node set $C(R)$ as the centres of sandboxes.

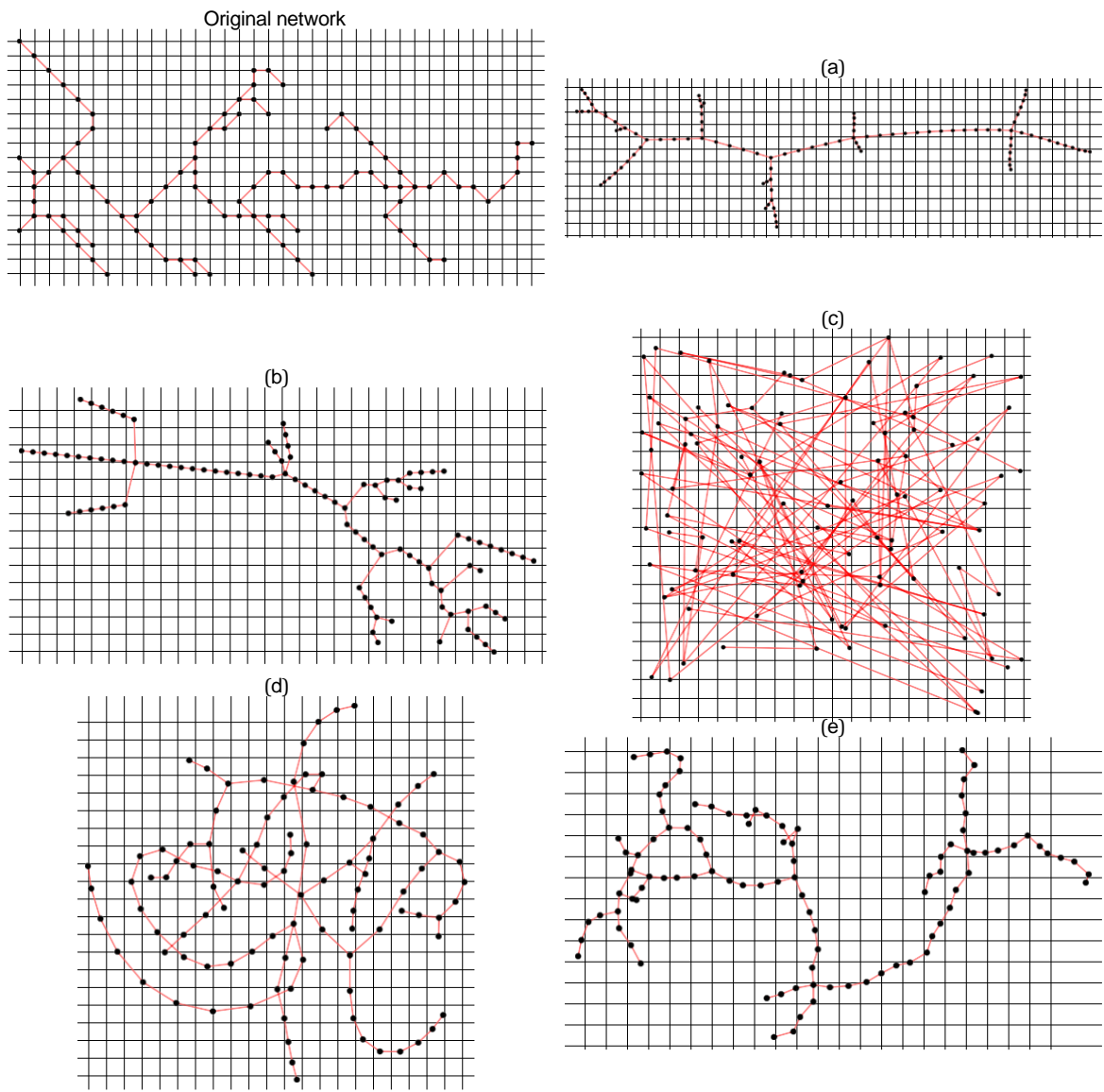

Figure 2: The same network in different geometrical representations where nodes have the same structure of connections and degree but are distributed differently in relation to each other in the plain. In each case, only the geometric coordinates of the location of each node change, and thus the length of the edge.

(3) Count the number of nodes $M_i(R)$ covered by the sandbox with center node $i \in C(R)$ and radius $R$ of the circle.

~~The number $M_i(R)$ includes nodes located in the area of the circle and forming a connected graph, i.e. in the circle there is a connection of each node with the central node $i$.~~

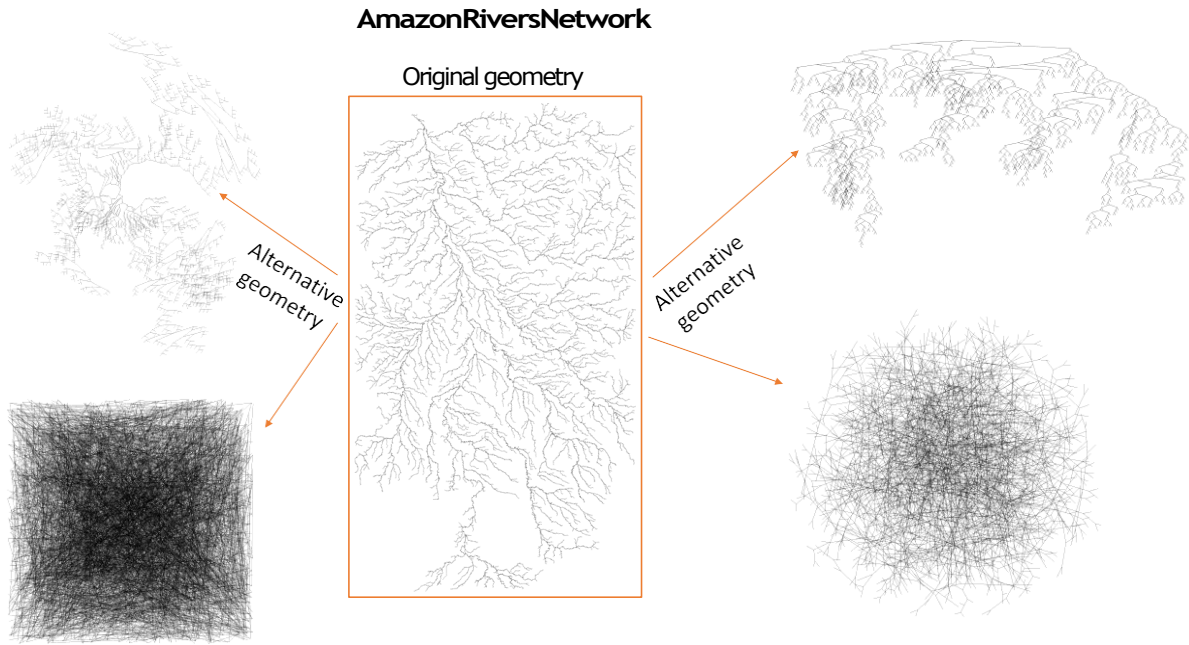

Figure 3: Alternative geometries of the real tributary network of the Amazon River.

The number $M_i(R)$ does not always include all nodes within the circle. As Fig.4 shows, the number $M_i(R)$ includes only those nodes that form a connected graph with node $i$ (nodes marked in black, although they are in a circle, are not included in $M_i(R)$).

Counting only nodes that form a graph connected to node $i$ ensures that we do not include nodes in a circle of radius $R$ that do not directly interact with the central node $i$ (we do not count nodes from which we cannot reach node $i$ along the edges). This approach to counting nodes is a convolution of the classic SB and the box-covering method: we count nodes from a given box (circles here, but there may also be another flat figure) but only those that have a network connection with the centre node. This process for $R = \{5.5, 9, 8.3\}$ is illustrated in Fig.4. The nodes included in the $M(R)$ number are marked in blue.

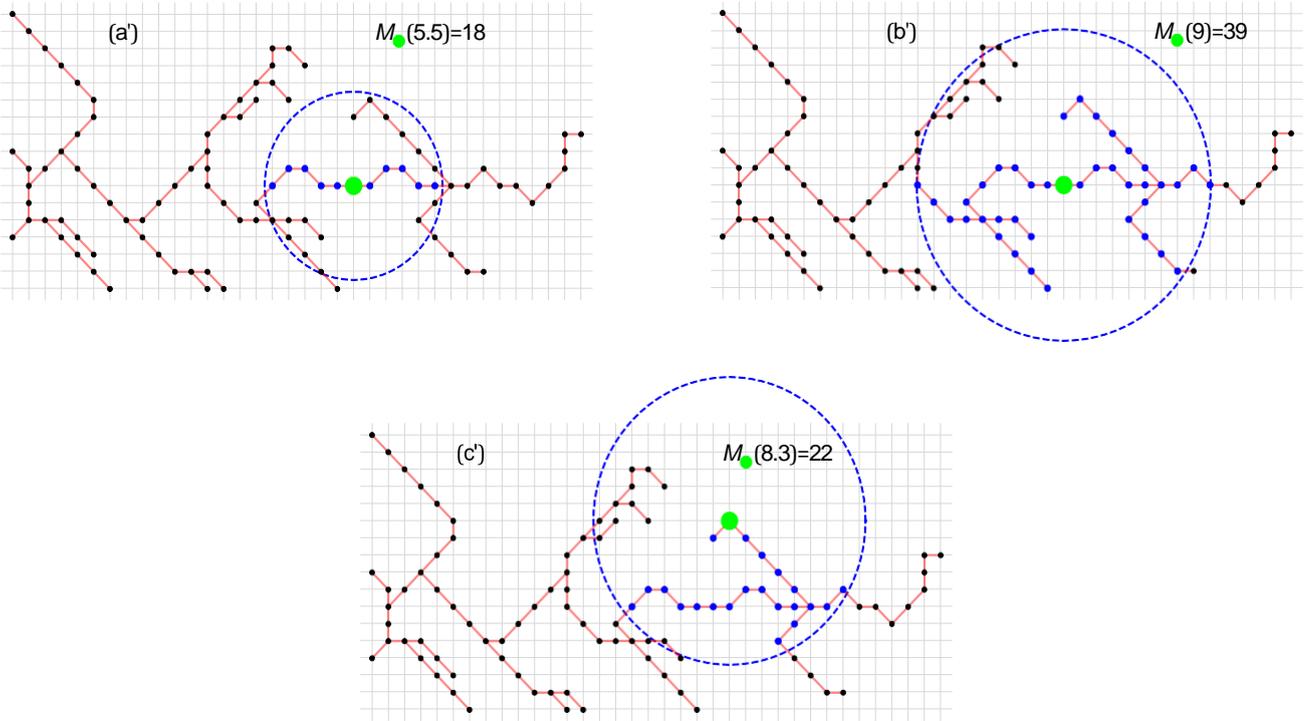

Figure 4: An example of counting network nodes in the GSB method for 3 radii $R$ of a circle. Examples (a$'$) and (b$'$) use two different circle radii $R = \{5.5, 9\}$ for the same central node. Example (c$'$) shows the counting of nodes for a different central node and radius $R = 8.3$.

(4) Steps (4)–(6) are the same as for the standard SB algorithm.

## 3 Numerical tests

In order to test the possible influence of the network geometry, i.e. the geometric position of the nodes in the network, let us consider three networks: the real network of the ridges of the Ligurian mountains, a tree-type network, and a scale-free network. For each type of network, multifractal analysis was performed for 4 alternative geometries of the same network. Changes in the geometry of each network were made in such a way that the location of the nodes was changed - thus both the distances between the nearest neighbors and the relative distances between all other nodes were changed (Fig.2) and (Fig.3 where alternative geometries of the Amazon river network are shown). It should be emphasised that changes in geometry do not change the degree of nodes, so the classic SB method, regardless of the geometry (appearance) of the network, gives the same result. Of course, there is a weighted variant of the classic SB that can take into account the distances between nodes [44], but unlike the GSB method proposed here, the weighted variant of the SB method cannot take into account the relative positions of all nodes. Geometric changes of the network (alternative geometries) were made in the Mathematica software system (the 'GraphM' [1] package was used for this). An example is shown in Fig.5 - there are 4 different geometric variants for Ligurian mountain ridges, tree-type network, and scale-free network, respectively [2].

For better figures clarity, the number of network nodes has been significantly reduced, which in the case of mountain ridges means that only a part of the mountain range is shown.

Fig.6, Fig.7, Fig.8 show the results of the multifractal analysis using the classic SB method and its modification proposed here, i.e., the GSB method. For the tree network (Fig.7) and the scale-free network (Fig.8), the network size is the same and is 40000 nodes. This number was chosen because the number of nodes for the real mountain ridge network is 40396 nodes (Fig.6). The results for generated (synthetic) tree and scale-free networks were averaged over 10 independent realisations.

The following notations are used: (a) are the results for the SB method; (b, c, d, e) are the results for the GSB. In the case of (a) and (b), a network with the same geometry is analysed. For easier comparison of results, the graphs are presented in the same numerical range. For each network, the fluctuation function $(< [M(R)]^{(q-1)} > /(q-1)) \equiv F$ was determined first (graphs with index (1)). Then, if the fluctuation function $F$ on the log-log scale is a straight line, we determine the fractal dimension $D_q$ and the singularity spectrum $f(\alpha)$ (graphs with index (2) and (3) respectively). Both in the case of the real network (Fig.6) and the other two synthetic networks (Fig.7, Fig.8), it can be seen that changing the network geometry changes its multifractal characteristics. In four cases, the fractal nature of the network was not found because the $F$ fluctuation function did not scale in the full range of $q$ values (Fig.6e) or there was no scaling in the entire $q$ range and $R$ radius (Fig.7e, Fig.8a, Fig.8e). Therefore, in these cases, it was not reasonable to determine subsequent multifractal characteristics.

The SB algorithm for the ridge network of the Ligurian mountains indicates the multifractal nature of the network (Fig.6 a1,a2,a3), where the width of the multifractal spectrum $\Delta \alpha = 0.19$.

Fig.6 (b1,b2,b3) shows the results of the GSB method for the original network geometry, which is geometrically similar to Fig.5(1). It can be seen that taking the geometry into account did not significantly affect the multifractal characteristics, i.e. $\Delta \alpha = 0.22$. We only observe a slight shift of the whole spectrum to the right. The next three lines of graphs in Fig.6 show the results of the GSB algorithm for 3 different geometries - similarly as in the first line in Fig.5 ((2),(3), (4)). It can be seen that the change in the real network geometry of the mountain ridge entails the change of multifractal characteristics, and the widest spectrum is observed for cases (c) and (d). In the case of (e), the fractal dimension $D_q$ and the multrifractal spectrum $f(\alpha)$ only for $q < 0$ is shown because only for them scaling in the full range of scales is observed. Furthermore, for this type of geometry (analogue of Fig.5 (4)), where the nodes are randomly distributed, the shift of the whole spectrum towards $\alpha = 1$ is visible, suggesting depletion of the complexity of the network, and elements of the monofractal nature of the network can only be seen for $q < 0$.

---

[1]'IGraphM' is a Mathematica package for use in complex networks and graph theory research. The creator of the interface is Szabolcs Horvát [48].

[2](1) IGLayoutFruchtermanReingold (the vertices are deployed on a plane according to the Fruchterman-Reingold algorithm) [49];

(2) IGLayoutReingoldTilfordCircular (the vertices are deployed on a plane according to the Reingold-Tilford algorithm) [50];

(3) IGLayoutGraphOpt (algorithm by Michael Schmuhl) [51];

(4) IGLayoutRandom (the vertices are deployed uniformly and randomly on a plane) [52].

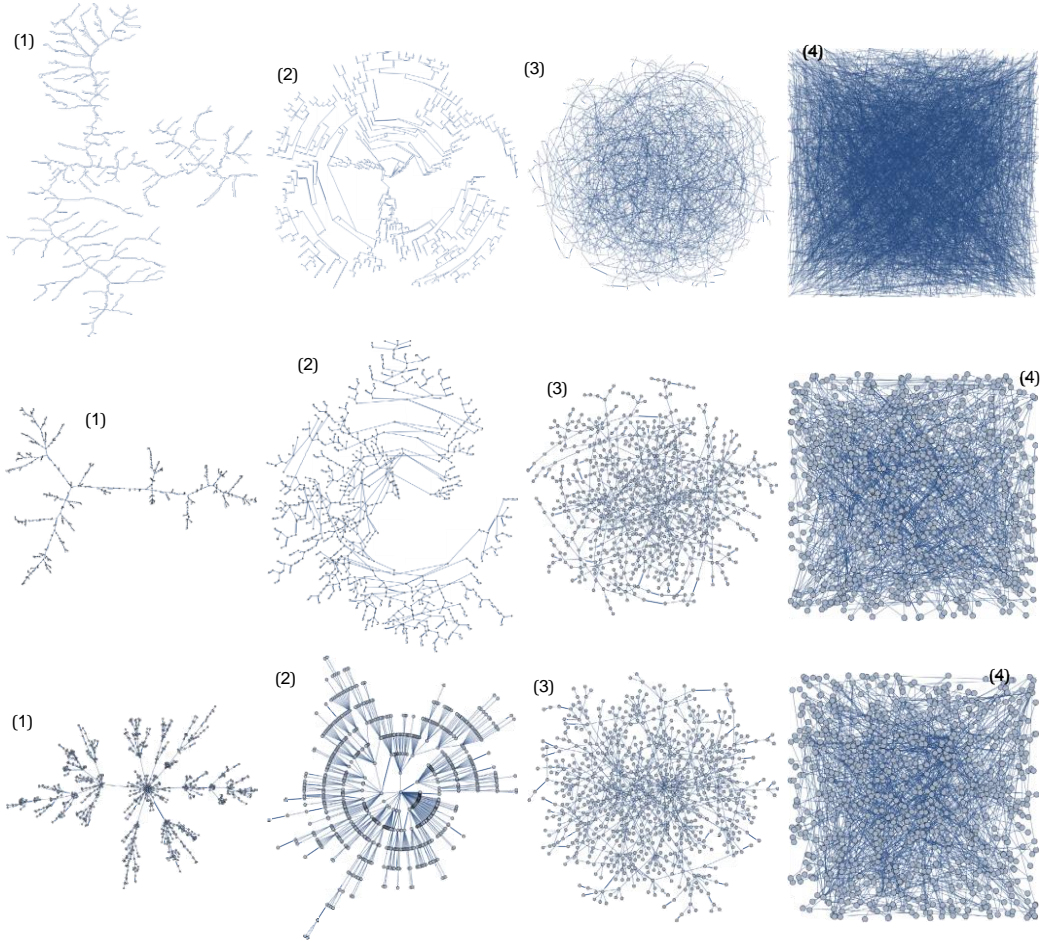

Figure 5: Four variants of geometry (1–4) for three types of networks: top - a network of Ligurian mountain ridges; middle - tree network; bottom - scale-free network. Changing the geometry consists in changing the position of the nodes on the plane, and does not change the nodes' degree. In order to better visualise networks with a small number of nodes are shown.

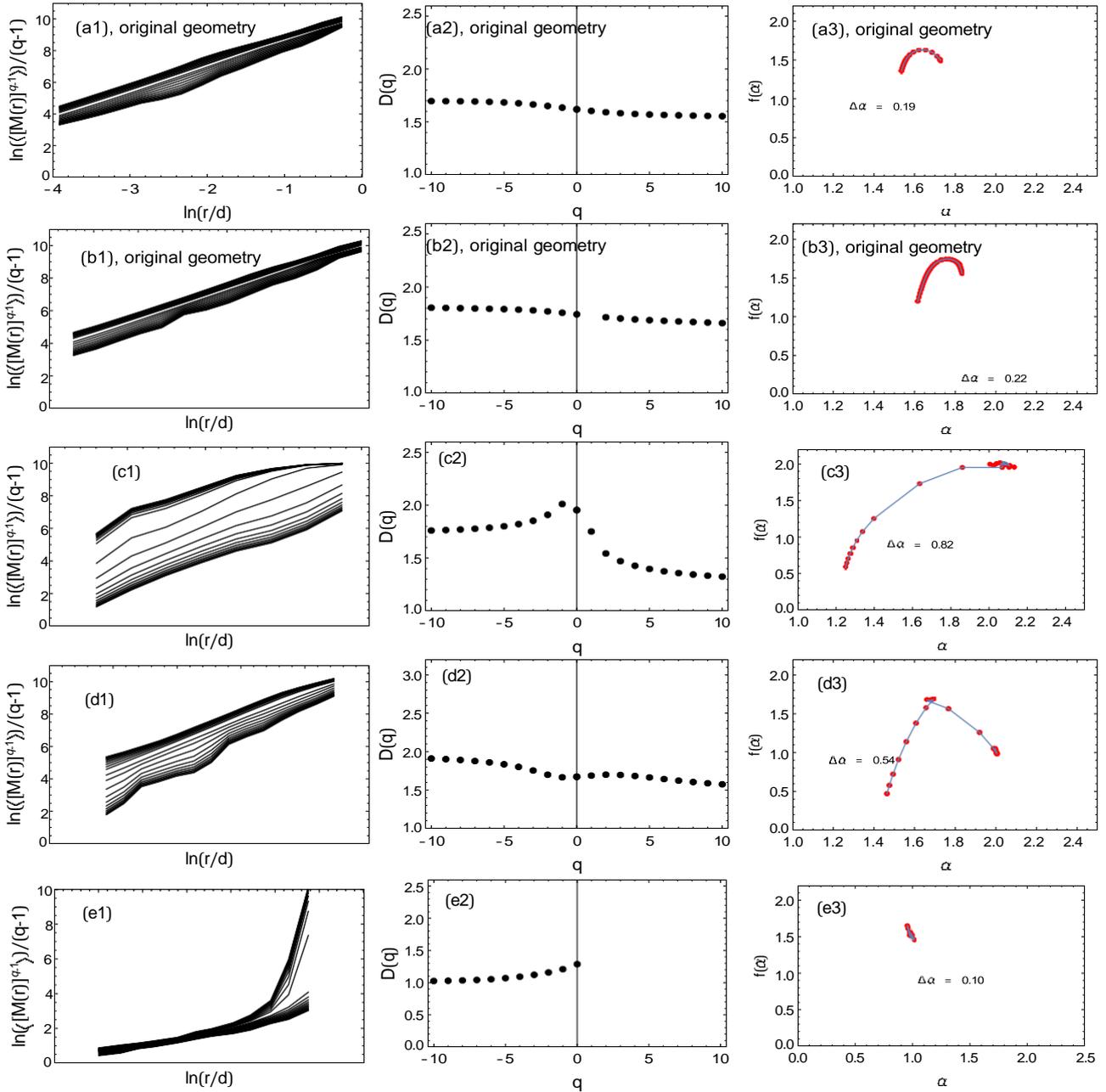

Figure 6: The multifractal analysis using the classic sandbox algorithm (a1, a2, a3) and its modification i.e. the Geometric sandbox method (other graphs) of the network of Ligurian mountain ridges. The first two rows are the results for the original real network; rows (c, d, and e) are the results for alternative geometries of the original network using the GSB algorithm. Columns (1), (2), (3) show fluctuation functions, fractal dimension, and multifractal spectrum, respectively.

Fig. 7 (a–e) shows the results for a synthetic tree network for geometric representations as in Fig. 5 (middle 1–4). The results of representation (1) are shown in a1,a2 a3 (SB), and b1,b2,b3 (GSB). It should be noted that example (1) is qualitatively similar to the real ridge geometry of the Ligurian mountains; therefore, the results are similar to the results for the original mountain geometry. For geometries (2) and (3) we can also see the influence of geometry on the fractal nature of the lattice (Fig.7 (c,d)). For geometry (2), the spectrum is the widest and most shifted to the right. In the case of (e), for geometry (4), where the nodes are arranged randomly, similarly to the real network, we observe the lack of scaling of the fluctuation function in the interval $r$ long enough to confirm the fractal nature of this network. Insignificant scaling elements can be seen only in a short range of small lengths $r$. Therefore, no attempt was made to estimate $D_q$ and $f(\alpha)$, thus claiming that this type of network with such geometry is not fractal.

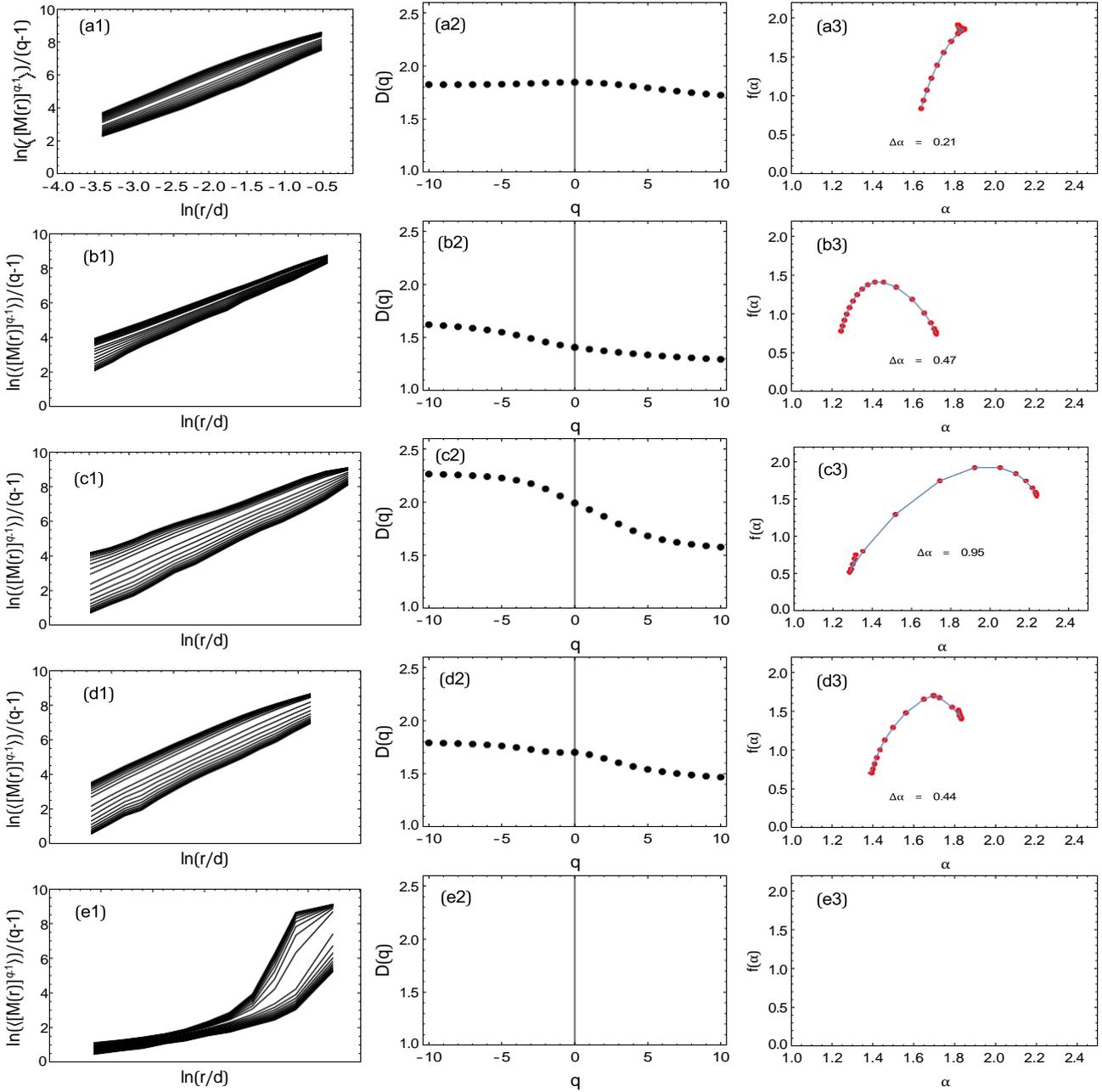

Figure 7: The multifractal analysis using the classic sandbox algorithm (a1, a2, a3) and its modification i.e. the geometric sandbox method (other graphs) of a synthetic tree-type network. In rows, are the results for alternative geometries of the network. Columns (1), (2), (3) show fluctuation functions, fractal dimension, and multifractal spectrum, respectively.

Fig.8 (a–e) shows the results for a synthetic Barabasi-Albert type scale-free network. In this type of network, the degree distribution has a power-low nature, which may suggest the fractality of this network. However, the classic SB method suggests quite the opposite (Fig.8 (a)) - the fluctuation function does not scale, so the fractal dimension $D_q$ and the multifractal spectrum $f(\alpha)$ do not exist. However, if we apply the GSB method taking into account the geometry of the network, fractality appears and, as in the previous examples, is dependent on the type of geometry (Fig.8 (b–d)). In cases (b,c,d) the $F$ fluctuation function is exponential in the whole range $r$ and the multifractal spectra are relatively wide ($\Delta\alpha$ takes values 0.81, 0.82, 1.3). Significantly higher than zero values of $\Delta\alpha$ testify to the complex nature of these networks. For geometry (e) where the nodes are distributed randomly (in the geometrical sense), similarly to the previously discussed types of network, the fluctuation function is not a power-low; therefore, it is not justified to calculate other fractal characteristics.

## 4 Conclusions

The article presents a different approach to the study of multifractality of complex networks. For this purpose, we proposed a modification of the sandbox method. The Geometric sandbox takes into account both connections between nodes and the location coordinates of nodes (network geometry). This approach combines both a typical sandbox network approach (only considers nodes connected to each other) and a box-counting approach (only nodes lying within a circle of radius R). We have tested our method for different geometrical variants of the same synthetic and real networks. It has been shown that networks with geometry that are pleasing to the human eye have wider multifractal spectra, in contrast to those where the nodes are randomly distributed (these types of network are not fractal). The test results confirm that the complexity of the network (its multifractal characteristics) is sensitive to changes in its geometry (coordinates of nodes).

Therefore, it seems that the method proposed here can be a useful tool for identifying and studying the degeneration of network multifractal structures not only in the case of 2D but also in 3D networks, where geometry often plays an important role, for example, the study of changes in DNA structures or brain neural networks.

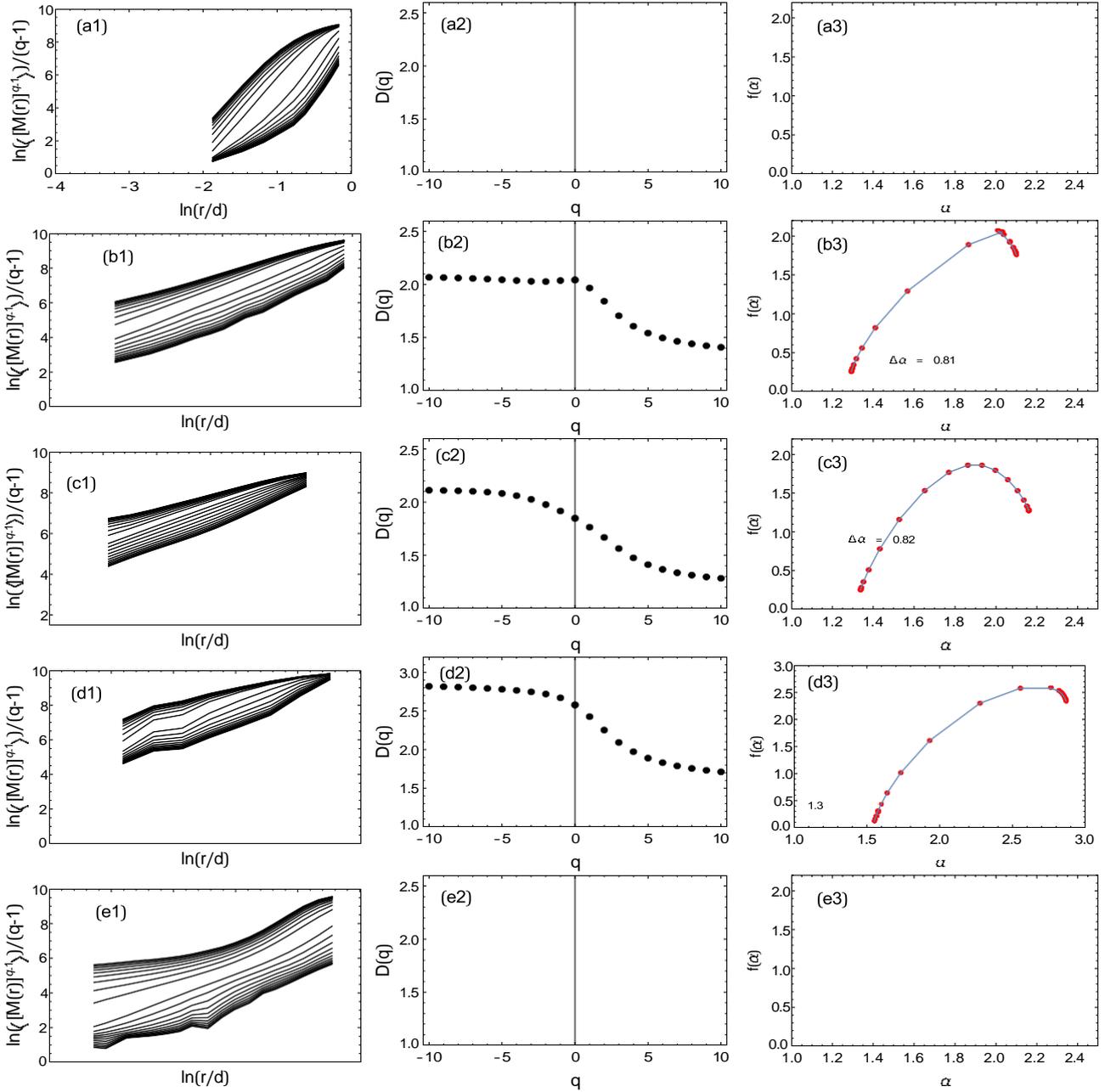

Figure 8: The multifractal analysis using the classic sandbox algorithm (a1, a2, a3) and its modification i.e. the geometric sandbox method (other graphs) of a synthetic scale-free network. In rows, are the results for alternative geometries of the network. Columns (1), (2), (3) show fluctuation functions, fractal dimension, and multifractal spectrum, respectively.